\documentclass[12pt,a4paper]{article}
\usepackage[utf8]{inputenc}
\usepackage[plainpages=false,pdfpagelabels=true]{hyperref}
\usepackage{amssymb,amsthm}
\usepackage[margin=1.5 in]{geometry}
\usepackage{graphicx}
\usepackage{a4wide}  
\usepackage{amsfonts}
\usepackage{amsmath}
\usepackage{bbm}
\usepackage{booktabs} 
\usepackage{multicol}
\usepackage{ifpdf}
\ifpdf

\hypersetup{%
  pdftitle    = {On massive particle surfaces, partial umbilicity and circular orbits}
  pdfkeywords = {Massive particle, surfaces, Jacobi, curvature},
  pdfauthor   = {Oscar},
  plainpages  = true,
  colorlinks  = true,
  citecolor   = blue,
  urlcolor    = blue,
  linkcolor   = black
}

\newcommand{\arxiv}[1]{{\tt
\href{http://www.arXiv.org/abs/#1}{#1}}}
\newcommand{\doi}[1]{{\tt DOI:\href{https://doi.org/#1}{#1}}}

\makeatletter
\@addtoreset{equation}{section}
\makeatother

\begin{document}

\begin{center}

{\Large {\bf On massive particle surfaces, partial umbilicity and circular orbits}}

\vspace{1.5cm}

\renewcommand{\thefootnote}{\alph{footnote}}
{\sl\large  Boris Berm\'{u}dez-C\'{a}rdenas\footnote{E-mail: bxbermudez [at] uc.cl } and Oscar Lasso Andino}\footnote{E-mail: {oscar.lasso [at] udla.edu.ec}}

\setcounter{footnote}{0}
\renewcommand{\thefootnote}{\arabic{footnote}}

\vspace{1.5cm}

{\it $^{a}$ Facultad de Matemáticas, Pontificia Universidad Católica de Chile, Avenida Vicuña Mackenna 4860, Santiago, Chile.}
\ \vspace{0.3cm}

{\it $^b$Escuela de Ciencias Físicas y Matemáticas,Universidad de Las Américas, Redondel del ciclista, Antigua vía a Nayón, C.P. 170504, Quito, Ecuador\\} \vspace{0.3cm}

\vspace{1.8cm}

{\bf Abstract}\\
\end{center}

The generalization of photon spheres by considering the trajectories of massive particles leads to the definition of Massive Particle Surfaces (MPS). These surfaces, built with the trajectories of massive particles, have a partial umbilicity property. Using the geodesic and Gaussian curvature of the Jacobi metric (a Riemannian metric), we derive a general condition for the existence of a Massive Particle Surface defined for an asymptotically flat spacetime metric. Our results can be applied to the worldlines of charged massive particle surfaces. We provide a simple characterization for timelike and null trajectories using a Riemannian geometric approach. We are able to recover the results for the existence of Light Rings (LR's) and timelike circular orbits (TCO's). We show how an event horizon gets characterized using the curvatures of a Riemannian metric. We discuss several examples, where we derive conditions for the existence of photon sphere and a massive particle surface. We calculate the radius of the photon sphere and the radius of the Innermost Stable Circular Orbits (ISCO).

\section{Introduction}

The study of astrophysical black holes has increased exponentially with the spectacular discovery of gravitational waves. The shadows, photon surfaces and other properties of black holes can be characterized geometrically. Particularly, the existence of Light Rings (LR's) or Timelike circular orbits (TCO's) can be studied by using Riemannian geometry. In the first case, the optical metric from a static spacetime together with the Haddamard theorem have been used to study photon orbits stability and the black hole shadows \cite{Qiao:2022jlu}, for a similar study using a more general metric  see \cite{Qiao:2022hfv}.  In the second case, the Jacobi metric has been used to study the trajectories of massive particles. \cite{Cunha:2022nyw}.  The Jacobi metric is a Riemannian metric obtained by projecting a Lorentzian metric over the direction of a Killing vector.\footnote{If the Killing vector is $\partial_t$, the projection is done over surfaces of constant energy.}. The Jacobi metric inherits the geodesic structure of the ambient spacetime metrics and has been calculated for different spacetimes \cite{Gibbons:2015qja,Arganaraz:2019fup,Duenas-Vidal:2022kcx,Das:2016opi,Maraner:2019ftj}.\\
The trajectories of massive particle surfaces can be studied from a purely geometric point of view. Using null geodesics, a photon surface can be built \cite{Claudel:2000yi,Virbhadra:2002ju}, and for static spherically symmetric spacetimes, this photon sphere is umbilic \cite{Senovilla:2011np,Okumura:1967}. A similar construction, but using the massive particle trajectories leads to a new concept, the Massive Particle Surfaces (MPS). The MPS are partially umbilic and have been mathematically defined recently \cite{Bogush:2023ojz} . In order to determine if a hypersurface is a MPS a \textit{master equation} has been defined \cite{Kobialko:2022uzj,Bogush:2024fqj}. If the right side of the master equation is constant, then the hypersurface in consideration is a MSP. This master  equation is directly related to the fact that the MPS has the property of being partially  umbilic. In this article, following a completely different approach, we found an equation that needs to be satisfied in order to have a MPS. The equation that we propose is built using the conformal factor and the component $g_{\phi\phi}$ of the Jacobi metric obtained by the projection of the spacetime metric over surfaces of constant energy.\footnote{Note that the MPS are timelike hypersurfaces defined over a Lorentzian manifold, but the Jacobi metric is a Riemannian metric.}.\\
Recently, a number of theorems regarding the existence of Light Rings (LR's) have been proved  \cite{Cunha:2020azh,Cunha:2017qtt,Ghosh:2021txu}. Similarly, in \cite{Cunha:2022nyw} the Timelike Circular Orbits (TCO's) have been studied by analyzing the Gaussian and geodesic curvature of the Jacobi metric. Part of this work represents a generalization of these results.\\
Our approach is simpler and it works also for MPS's built with the worldlines of charged particles. We start by calculating the Jacobi metric, and then by enforcing the vanishing of the  geodesic curvature we obtain an equation involving the first derivative of the conformal factor of the Jacobi metric. From this new equation the master equation for the existence of a MPS is derived showing a direct relationship between massive particle surfaces and the Jacobi metric.\\
The continuum surface made of TCO's is characterized by the energy and angular momentum of the particle. On the other side, the Jacobi metric encodes properties of the particle: mass, charge and momentum. Therefore, it is expected that the information related to TCO's is codified in the Jacobi metric. The characterization that we present is general enough for reproducing the results concerning the stability of LR's and  TCO's. The Innermost Stable Circular Orbit (ISCO) is a marginally stable circular orbit\footnote{The marginally stable circular orbits MSCO's constitute the boundary between regions of different stability.} and divides the TCO's surface between regions with different stability. The information of ISCO's and of the photon sphere is embedded in the Jacobi metric and we will study them in full generality. \\
Using the geodesic and Gaussian curvatures of the Jacobi metric we study the near horizon geometry of black holes. We are able to see how the black hole event horizon, a Lorentzian property, is characterized in a Riemannian geometry.\\
In section \ref{mps:1} we present a review of the results regarding massive particle surfaces, where a subsection on the partial umbilicity property has been detailed. We only present the neutral particle surface results, although our results can be applied to charged particles, see appendix \ref{app:1}. In section \ref{jmic:1} we calculate the Gaussian and geodesic curvatures of the Jacobi metric. We study its near horizon and asymptotic limit behaviour and we recover the null cases. In section   \ref{slt:1} we use the Gaussian and geodesic curvatures of the Jacobi metric for studying the stability of LR's and TMCO's. In section \ref{sec:5} we derive the condition for the existence of the MPS and characterize partial umbilicity of the MPS. Finally, in section  \ref{discussion} we present the discussion section. In appendix \ref{app:1} we describe the massive particle surface master equation for charged particles.

\section{Massive particle surfaces}\label{mps:1}

Here we present a brief review of the massive particle surfaces, see \cite{Bogush:2023ojz,Kobialko:2022uzj} for the whole theory and more examples. We will focus on the case without electrical charge but our results can be generalized easily to the charged case, for the charged massive particle surface see Appendix \ref{app:1}. 
We start by defining a Lorentzian manifold $\mathcal{M}$ on dimensions $d\geq 4$ and with metric tensor $g_{\mu\nu}$. The worldline $ x^{\mu}$ of a test particle of mass $m$ in the manifold $\mathcal{M}$ is described by 
\begin{equation}
v^{\mu}\nabla_{\mu}v^{\nu}=0,\,\,\,\,\,\,v^{\mu}v_{\mu}=-m^2,
\end{equation} 
where $v^{\mu}=\frac{dx^{\mu}}{ds}$ is the four-velocity of the particle, and $s$ is an affine parameter. We will work with metrics such that the Lie derivative satisfies
\begin{equation}
\mathcal{L}_{k}g_{\mu\nu}=\nabla_{(\mu}k_{\nu)}=0,
\end{equation}

where $k_{\mu}$ is a timelike Killing vector. The previous assumption is satisfied by static and stationary spacetimes\footnote{This assumption includes spacetimes non-necesary asymptotically flat but we will consider only those.} Moreover, the total energy of the particle $\mathcal{E}=k_{\mu}v^{\mu}$ is conserved, indeed, it is straightforward  to show that 
\begin{equation}
\frac{d\mathcal{E}}{ds}=-v^{\mu}v^{\nu}\nabla_{\mu}k_{\nu}=0.
\end{equation}
The vectors $v^{\mu}$ constitute a basis for all worldline particles with fixed energy $\mathcal{E}$ and mass $m$ moving through a point of spacetime. These worldlines define a timelike surface $\mathcal{S}$ of dimension $d-1$. Let us consider a unit normal vector $n^{\mu}$, then the induced hypersurface metric can be written:
\begin{equation}
h_{\mu\nu}=g_{\mu\nu}-n_{\mu}n_{\nu}.
\end{equation}
Moreover, we can define
\begin{equation}
h^{\mu}_{\nu}=\delta^{\mu}_{\nu}-n^{\mu}n_{\nu},
\end{equation}
therefore the second fundamental form can be written 
\begin{equation}
\Pi_{\mu\nu}=h^{\alpha}_{\mu}h^{\beta}_{\nu}\nabla_{\alpha}n_{\beta}.
\end{equation}

The Killing vector $k^{\mu}$ can be projected over the hypersurface $\mathcal{S}$
\begin{equation}
k^{\mu}=\tilde{k}^{\mu}+k_{\bot}n^{\mu},\,\,\,\,\,\tilde{k}^{\mu}n_{\mu}=0.
\end{equation}
If $\tilde{k}_{\mu}\tilde{k}^{\mu}\neq 0$ a vector $v^{\mu}$ can be decomposed as: 
\begin{eqnarray}
v^{\mu}&=&(\mathcal{E}/\tilde{k}^2)\tilde{k}^{\mu}+u^{\mu},\\ \tilde{k}^{\mu}u^{\mu}&=&n_{\mu}u^{\mu}=0,\,\,\,\, u^2=-m^2-\mathcal{E}^2/\tilde{k^2}.
\end{eqnarray}
Here, $u^{\mu}$ is a tangent vector to $\mathcal{S}$ and orthogonal to $\tilde{k}^\mu$. For MPS a condition that restricts the energy of the trajectories is given by
\begin{equation}\label{rest:1}
0<|\tilde{k}^{\mu}\tilde{k}_{\mu}|\leq \mathcal{E}^2/m^2
\end{equation} 
This inequality is a manifestation of the restrictive motion over massive particle surfaces. The inequality determines the zones over $\mathcal{S}$ that are allowed for motion. Evidently, because $m^2=0$ this restriction is not present in the massless case and the restriction \eqref{rest:1} leads to the conformal invariance of the null geodesic equations.\\
Here we present the formal definition of a massive particle surface as it is given in \cite{Bogush:2023ojz,Kobialko:2022uzj}.  \\

\textbf{Definition} \cite{Kobialko:2022uzj}  Let $M$ be a Lorentzian manifold of dimension $d\geq 4$. A massive particle surface $S_{\mathcal{E}}$ is an immersed, timelike, nowhere orthogonal to Killing vector $k^{\mu}$, such that for every point $p\in S_{\mathcal{E}}$ and every vector $v^{\mu}_{p}\in T_{p}S_{\mathcal{E}}$ such that $v^{\mu}k_{\mu}\vert _p=-\mathcal{E}_{k}\vert_{p}$ with $v^{\mu}v_{\mu}\vert_{p}=-m^2$, there exists a worldline $x^{\mu}$on $\mathcal{M}$ for a particle of mass $m$ and total energy $\mathcal{E}$. Moreover, $v^{\mu}(0)=v^{\mu}\vert_{p}$ and $x^{\mu} \subset S_{\mathcal{E}}$.\\
The previous definition reduces to the definition of photon surfaces if $m=0$. Note that massive particles do not posses conformal invariance and therefore it is necessary to fix the energy.

\subsection{Partially Umbilic surfaces}

It is well known that a photon surface is timelike totally umbilllic \cite{Claudel:2000yi,Senovilla:2011np}. This property can be stated mathematically by afirming that the second fundamental form $\Pi_{\mu\nu}$ is proportional to the induced metric $h_{\mu\nu}$. This property will help to study geodesic motion by using the geometric properties of the surface instead of solving the geodesic equations. This property is known to be satified by the photon surfaces of sphericallly symmetric spacetimes, such as Schwarzschild spacetime. However, photon surfaces of axi-symmetric spacetimes metrics, such as Kerr, do satisfy a weaker condition called partial umbilicity. Thus, the proportionality between the second fundamental form and the induced metric is impossed only over a subset of the tangent space.  On the other side, MPS are not a conformally invariant and not totally umbilic hypersurfaces \cite{Bogush:2023ojz,Bogush:2024fqj,Kobialko:2022uzj}, instead, a new partially umbilic condition is known to be satisfied\footnote{This partial umbilic property is the same partial umbilic condition satisfied by the photon sphere of the Kerr metric.}. Let us consider the set of basis vectors built with $\tilde{k}^{\alpha}$ and a set of linearly independent $n-2$ vectors  $\tau_{(i)}^{\alpha}$ defined in $S_{\mathcal{E}}$  such that
\begin{equation}
    \tau_{(i)}^{\alpha}\tilde{k}_{\alpha}=0.
\end{equation}
The principal curvatures $\lambda_{\tau_{i}}, \lambda_\kappa$  along the directions $\tau_{(i)}^{\alpha}$ and $\tilde{\kappa}^{\alpha}$ are \cite{Kobialko:2022uzj}:
\begin{equation}\label{pc:1}
\lambda_{\tau(i)}=\frac{\chi_\tau}{n-2},\,\,\,\,\,\,\,\lambda_\kappa=\lambda_{\tau(i)}(1+\frac{m^2}{\mathcal{E}}\kappa^2).    
\end{equation}
where
\begin{eqnarray}
\chi_{\tau}=\frac{n-2}{H}\chi_{\alpha}^{\alpha},\,\,\,\,\, \, \chi_{\alpha\beta}=\frac{\chi_{\tau}}{n-2}H_{\alpha\beta}, \\
H=H^{\alpha}_{\alpha},\,\,\,\,\, \,H_{\alpha\beta}=h_{\alpha\beta}+\frac{m^2}{\mathcal{E}^{2}}\tilde{\kappa}_\alpha\tilde{\kappa}_\beta.
\end{eqnarray}
The fact that the principal curvatures defined in \eqref{pc:1} are not equal implies that the surface $S_\mathcal{E}$ is not totally umbilic. In order to prove that a surface is a MPS the geometric condition of partial umbilicity has been translated to a master equation that needs to be satisfied \cite{Kobialko:2022uzj}, which for the neutral particle of mass $m$ can be written:
\begin{equation}\label{mastereq:1}
\mathcal{E}=\pm m \sqrt{\frac{\kappa^2 \chi_{\tau}}{W}},
\end{equation}
where 
\begin{equation}
\kappa^2=\tilde{\kappa}^{\mu}\tilde{\kappa}_{\mu},\,\,\,\,\,\,\,\,W=-\chi_{\tau}-(n-2)\tilde{\kappa}^{-2}\tilde{\kappa}^{\alpha}n^{\beta}\nabla_{\alpha}\tilde{\kappa}_{\beta}.
\end{equation}
If the right-hand side of the equation \eqref{mastereq:1} is constant, then the surface is a MPS. Note that $\mathcal{E}$ represents not only the energy, it can be any conserved quantity related with a Killing vector of the metric. Thus, if the Killing vector $\kappa^{\alpha}\partial_{\alpha}=\partial_{\phi}$ is considered, then the conserved quantity corresponds to the angular momenta $L$. A similar master identity needs to be satisfied when a charged MPS is considered, see Appendix \ref{app:1}.\\
Let us consider the asymptotically flat spacetimes of the form 
\begin{equation}\label{gmetr:1}
ds^2=-f(r)dt^2+\frac{1}{g(r)}dr^2+h(r)(d\theta^2+ \sin^{2}(\theta)d\phi^2).
\end{equation}

Asymptotic flatness enforce the conditions over the functions $f, g$ and $h$ in such a way that at infinity the Minkowski spacetime is recovered. Therefore,
\begin{equation}\label{asympc:1}
\lim_{r\rightarrow \infty}f(r)=1,\,\,\,\,\lim_{r\rightarrow \infty}g(r)=1,\,\,\,\,\lim_{r\rightarrow \infty}h(r)=r^2.
\end{equation}
The metric \eqref{gmetr:1} is a Lorentzian metric, and we are going to study its null and timelike geodesics using a Riemannian metric that encodes its geodesic structure. We are going to build a MPS and study its geometry.\\
A Killing vector of the metric \eqref{gmetr:1} is $k^{\alpha}\partial_{\alpha}=\partial_t$, then the master equation \eqref{mastereq:1} becomes
\begin{equation}\label{massivecond}
\frac{\mathcal{E}^2}{m^2}=\frac{f \partial_{r}\ln h}{\partial_r \ln \left(h/f\right)}=\frac{h' f^2}{h'f-hf'},\,\,\,\,\,\,\partial_{r}\left(\frac{h}{h \sin^2(\theta)}\right)=0.
\end{equation}

In the massless case the umbilicity condition reads
\begin{equation}
\partial_r \ln f=\partial r \ln h=\partial_r\ln (h \sin^2(\theta)),
\end{equation}
which leads to \cite{Claudel:2000yi} 
\begin{equation}\label{photoncond}
f'h-h'f=0.
\end{equation}
Here we can see that a purely geometric condition such as the umbilicity (partial umbilicity) of a massless (massive) surface leads to an equation that has a physical meaning, namely the existence of circular null/timelike geodesics and its stability. Indeed, the ISCO's can be obtained from the equation

\begin{equation}\label{isco:1}
\frac{d\mathcal{E}}{dr}=0.
\end{equation}
If the ISCO radius does not coincide with the radius of the photon surface and, in order to avoid horizons or singularities  $f\neq 0$ everywhere, then the condition \eqref{isco:1} can be written\footnote{Note also that $\mathcal{E}\neq 0$, it is because the trajectories are null or timelike}:
\begin{equation}\label{isco:2}
\left(2h'-\frac{h h''}{h'}\right)ff'-2hf'^2+hf f''=0.
\end{equation}

When $h=r^2$ we recover a very well known condition \cite{Claudel:2000yi} for the existence of ISCO's 
\begin{equation}\label{isco:3}
3f f'+r f f''-2rf'^2=0.
\end{equation}

If the metric has a horizon or a singularity the geometry of the MPS will change and it is not clear what happens around these points. However, as we are going to show, the curvatures of the Jacobi metric allows us to study these limits. This is one of the advantages of our approach, we can reach a singularity.

If there is a point $r_{H}$ on which $g(r_H)=0$  we say that $r_{H}$ is  a horizon of the metric \eqref{gmetr:1}. For the particular case of static black holes, the event horizon is also a Killing horizon \footnote{For static black holes the  horizon is a Killing horizon of the time traslation Killing vector $\partial_t$.} It is important to note that the existence of a horizon is not related whatsoever with the metric \eqref{gmetr:1} being a solution of the equations of motion of a given theory. The existence of a horizon is a geometric definition over a Lorentzian manifold, and we are going to see how it is characterized over a Riemannian  manifold.

Furthemore, we will focus on black holes whose surface gravity is different from zero \footnote{These black holes are called non-extremal and in that case the surface gravity is positive definite}. We will concentrate in studying the geometric properties of the metric  \eqref{gmetr:1} by projecting over surfaces of constant energy and constant momentum, the resulting Riemannin metric is known as the Jacobi metric, which in the massless case reduces to an optical metric.  \\

\section{The Jacobi metric and its curvatures}\label{jmic:1}

In this section we present some results about the Gaussian and  geodesic curvatures of the Jacobi metric obtained from the metric \eqref{gmetr:1}. We are able to study the MPS using the curvatures of a two dimensional Riemannian metric: the Jacobi metric \cite{Gibbons:2015qja,Arganaraz:2021fwu}.  In all cases, recovering the results for the photon sphere is done by setting $m=0$.

\subsection{The Jacobi metric}
The action $S=-m\int \mathcal{L}d\lambda $ for a massive particle moving in the spacetime described by the metric \eqref{gmetr:1} is defined using the Lagrangian $\mathcal{L}$
\begin{equation}
    \mathcal{L}=-m\sqrt{f(r)-g_{ij}dx^{i}dx^{j}},
\end{equation}
where $\lambda$ is a parameter along the geodesics. In all cases we will consider that the geodesics are parametrized by the coordinate time $t$.
The Hamiltonian of the sistem is obtained by a Legendre transform:
\begin{equation}\label{ham:1}
\begin{split}
    \mathcal{H}&=p_{i}\dot{x}^i-m\sqrt{f(r)-g_{ij}d\dot{x}^{i}d\dot{x}^{j}}\\
   & =\sqrt{f(r)^2(m^2+g^{ij}\partial_{i}S\partial_{j} S)},
    \end{split}
\end{equation}
where
\begin{equation}
p_{i}=\frac{\partial \mathcal{L}}{\partial \dot{x}^{i}}=\partial_i S.
\end{equation}
The Hamiltonian \eqref{ham:1} is independent of time, then the Hamilton equation leads to
\begin{equation}
   - \frac{\partial S}{\partial t}=\mathcal{H}=\mathcal{E},
\end{equation}
hence 
\begin{equation}\label{jacd:1}
\frac{f(r)^2}{\mathcal{E}^2-m^2 f(r)^2}g^{ij}\partial_i S\partial_j S=1.
\end{equation}
The equation \eqref{jacd:1} is the Hamilton equation for the geodesics of the metric:
\begin{eqnarray}\label{gJacobim}
J_{ij}dx^{i}dx^{j}=\left[\frac{\mathcal{E}^2-m^2f(r)}{f(r)}\right]\left(\frac{dr^2}{g(r)}+h(r)d\Omega^2\right),
\end{eqnarray}
where $d\Omega^2=(d\theta^2+\sin^2(\theta)d\phi^2)$. The metric \eqref{gJacobim} is a Riemannian metric known as the Jacobi metric \cite{Gibbons:2015qja,Arganaraz:2021fwu}. Geometrically, the Jacobi metric describes the geodesic motion when restricted to surfaces of constant energy $\mathcal{E}$.
For the photon trajectories  $(m=0)$ the metric \eqref{gJacobim} transforms to:

\begin{equation}\label{gJacobimop}
J_{ij}dx^{i}dx^{j}=\frac{E^2}{f(r)}\left(\frac{dr^2}{g(r)}+h(r)(d\theta^2+\sin^2(\theta)d\phi^2)\right).
\end{equation}
which can be written 
\begin{equation}
J_{ij}dx^{i}dx^{j}=E^2g_{ij}^{OP}dx^{i}dx^{j},
\end{equation}
where $g_{ij}^{OP}$ is the optical metric.The optical metric is obtained by imposing the condition $ds^2=0$ on the spacetime metric and solving for $dt$. The geodesics of the optical metric are light rights that are spatial projections of the null geodesics of the original spacetime metric. Thus, the optical metric obtained from \eqref{gmetr:1} is given by
\begin{equation}\label{Jacobimop1}
\tilde{g}_{ij}^{OP}dx^{i}dx^{j}=\frac{1}{f(r)}\left(\frac{dr^2}{g(r)}+h(r)d\phi^2\right),
\end{equation}
where, because of spherical symmetry, we have set $\theta=\frac{\pi}{2}$. The metric \eqref{Jacobimop1} is a $2-$dimensional Riemannian metric, and the theorems of Riemannian geometry can be used to study its properties. Using the Hadammard theorem\footnote{The Hadammard theorem states that for any two-dimensional Riemannian manifold with non-positive Gaussian curvature two arbitrary points are connected by a segment, which is unique if the manifold is simply connected, but if the manifold is not connected there is a unique geodesic joining the two points for each homotopy class and this geodesic curve minimizes the length in this homotopy class \cite{Berg:1988}.}  a criteria for the stability of photon orbits was found in \cite{Qiao:2022jlu}. The result was proven for static spacetimes such that $f(r)=g(r)$ and $h(r)=r^2$. Due to the fact that the Jacobi metric \eqref{gJacobimop} is proportional to $g_{ij}^{OP}$ the results are the same for the metric \eqref{gJacobimop}, see also \cite{Qiao:2022hfv} for more examples.\\ 
In order to study the properties of the massive particle surfaces/photon spheres we are going to use the Jacobi metric \eqref{gJacobimop} which we  write as
\begin{equation}\label{jacF:1}
J_{ij}dx^{i}dx^{j}=F(r)\left(\frac{dr^2}{g(r)}+h(r)d\phi^2\right),
\end{equation} 
where 
\begin{equation}\label{Fjac}
F(r)=\frac{\mathcal{E}^2-m^2f(r)}{f(r)}.
\end{equation}

In the next sections we are going to calculate the Gaussian and geodesic curvature of the Jacobi metric \eqref{jacF:1} and study their  properties.

\subsection{Gaussian curvature}

The Gaussian curvature of a $2-$dimensional metric of the type
\begin{equation}\label{2dg}
ds^2=Ddu^2+Gdv^2,
\end{equation}
can be calculated by \cite{Kre:1991}
\begin{equation}\label{gaussg}
\mathcal{K}=-\frac{1}{2\sqrt{DG}}\left(\frac{\partial}{\partial u}\left(\frac{\partial_{u}G}{\sqrt{DG}}\right)+\frac{\partial}{\partial v}\left(\frac{\partial_{v}D}{\sqrt{DG}}\right)\right).
\end{equation}

By setting $D=\frac{F(r)}{g(r)}$ and $G=F(r)h(r)$  in \eqref{gaussg} the Gaussian curvature of \eqref{jacF:1} can be calculated:
\begin{equation}
\begin{split}
\mathcal{K}&=-\frac{1}{4h^2F^3}\left(g'\left(h^2FF'+F^2hh'\right)
\right.\\
&
\left.
+g \left(FhF' h'+2 F^2 h h''-2 h^2 \left(F'^2-F F''\right)-F^2 h'^2\right)\right)\label{gGaussianF}.
\end{split}
\end{equation}

Replacing \eqref{Fjac} in \eqref{gGaussianF} and after rearranging terms we find that:

\begin{equation}
\begin{split}
\mathcal{K}&=-\frac{\mathcal{E}^2}{\left(\mathcal{E}^2-m^2f\right)^2}\left[g'\left(\frac{h'f-hf'-\frac{m^{2}}{\mathcal{E}^2}f^2h'}{4h}\right)
\right.\\
&
\left.
+\frac{g}{2}\left(f'\left(\frac{2f'}{f}-\frac{h'}{2h}\right)-f''-\left(\frac{(h')^2}{2h^2}-\frac{h''}{h}\right)f
-\frac{(f')^2}{\left(1-\frac{m^2}{E^2}f\right)f}+\frac{m^2}{E^2}\left(\frac{(h')^2}{2h^2}-\frac{h''}{h}\right)f\right)\right]\label{gaussm:1}.
\end{split}
\end{equation}

When $h(r)=r^2$ we obtain
\begin{equation}
\begin{split}
\mathcal{K}&=-\frac{1}{\mathcal{E}^2\left(1-\frac{m^2}{\mathcal{E}^2}f\right)^2}\left[g'\left(\frac{2f-rf'-2\frac{m^{2}}{\mathcal{E}^2}f^2}{4r}\right)
\right.\\
&
\left.
+\frac{g}{2}\left(f'\left(\frac{2f'}{f}-\frac{1}{r}\right)-f''-\frac{(f')^2}{\left(1-\frac{m^2}{\mathcal{E}^2}f\right)f}\right)\right]\label{gaussm:2}.
\end{split}
\end{equation}

If we also set $m=0$ in \eqref{gaussm:2}  we recover the results for the LR's obtained in \cite{Qiao:2022jlu,Cunha:2022nyw}.

For asymptotically flat spacetimes we impose conditions \eqref{asympc:1},  then the Gaussian curvature \eqref{gaussm:1} satisfies 
\begin{equation}
\lim_{r\rightarrow \infty}\mathcal{K}=0
\end{equation}
The limit of the Gaussian curvature was expected. In the asymptotic region the spacetime is flat and therefore its intrinsic curvature vanishes. The Jacobi metric is asymptotically euclidean. 

If we set $h(r)=r^2$ an replace expression \eqref{Fjac} in \eqref{gaussm:1} we obtain

\begin{equation}\label{kcirc}
K_{circ}=-\frac{g}{2\mathcal{E}^2 r\left(1-\frac{m^2}{\mathcal{E}^2}f\right)^2}\left(r (f')^2\left(\frac{1-\frac{2m^2}{E^2}f}{1-\frac{m^2}{\mathcal{E}^2}f}\right)-f (f'+r f'')\right)
\end{equation}

Expression \eqref{kcirc} works also for the null case only by setting $m=0$. This is a generalization of previous results in the literature \cite{Cunha:2020azh}. Our approach let us to study the massive case always carrying the null case. Moreover, note that $K_{circ}$ depends on $\mathcal{E}$, the energy of the trajectory, which is a constant of motion. 

\subsubsection{Near horizon geometry}

When we move towards the horizon $r_{H}$ the Gaussian curvature \eqref{gGaussianF} behaves as 
\begin{align}
\lim_{r\rightarrow r_h}\mathcal{K}&=\frac{1}{E^2}\lim_{r\rightarrow r_h}\left[ \frac{g' f'}{4}\right],
\end{align}
This result shows, as expected, that the Gaussian curvature is not affected by the mass terms. It only depends on the derivative of the spacetime metric functions $f'$ and $g'$. Note that for the cases when $f(r)=g(r)$ the expression simplifies to
\begin{align}
\lim_{r\rightarrow r_h}\mathcal{K}&=\frac{1}{4E^2}\lim_{r\rightarrow r_h}f'^2.
\end{align}
The previous relation implies that for metrics satisfying $f(r)=g(r)$ the Gaussian curvature is positive at the horizon. Due to the fact that at infinity the Gaussian curvature vanishes then it has to decrease from positive values until it reaches  zero at infinity, no matter what happens in between. A simple example is given by the Shchwarzschild metric which satisfies:
\begin{align}
\lim_{r\rightarrow r_h}\mathcal{K}_S&=\frac{1}{4E^2}\lim_{r\rightarrow r_h}\frac{M^2}{r_{H}^4}=\frac{1}{16E^2M^2}.
\end{align}
If the mass of the black hole is big then the Gaussian curvature of the horizon is small. This results is in accordance with what is know about surfaces. For a fixed energy, the surface associated to the event horizon has intrinsic curvature, and  behaves as the Gaussian curvature of a sphere whose radius is $M$.

\subsection{Geodesic curvature}

Another curvature measure is the geodesic curvature. The geodesic curvature $\kappa_{g}$ measures how far  a curve is of being a geodesic.  For a $2-$dimensional metric of the form \eqref{2dg} the geodesic curvature can be calculated using
\begin{equation}
\kappa_{g}=\frac{1}{2\sqrt{D}}\frac{\partial\ln(G)}{\partial r}\vert_{r=r_{o}}.
\end{equation}
For metric \eqref{jacF:1} the geodesic curvature becomes
\begin{equation}\label{geoF:3}
\kappa_{g}=\frac{1}{2}\sqrt{\frac{g}{F}}\left(\frac{F'}{F}+\frac{h'}{h}\right).
\end{equation}
This curvature is an intrinsic quantity when calculated in the $2-$dimensional surface that is defined by \eqref{jacF:1}. Replacing \eqref{Fjac}  in \eqref{geoF:3} we obtain
\begin{equation}\label{gaussianc:1}
\kappa_{g}=\frac{\sqrt{g(r)}}{E\left(1-\frac{m^2}{E^2}f(r)\right)^{3/2}\sqrt{f(r)}}\left(\frac{h'f-hf'-\frac{m^2}{E^2}h'f^2}{2h}\right).
\end{equation}
For photon orbits ($m=0$) the previous expression transforms to
\begin{equation}\label{gaussianc:2}
\kappa_{g}=\frac{\sqrt{g(r)}}{E\sqrt{f(r)}}\left(\frac{h'f-hf'}{2h}\right),
\end{equation}
if we also take $h(r)=r^2$, in the geodesic curvature \eqref{gaussianc:2} we recover the result found in \cite{Qiao:2022jlu,Cunha:2020azh}, namely $\kappa_{g}=\frac{\sqrt{g(r)}}{E\sqrt{f(r)}}\left(\frac{2f-rf'}{2r}\right).$

As we did with the Gaussian curvature, we would like to know what happens at infinity with $\kappa_g$. Thus, the expression in \eqref{gaussianc:1}   behaves at infinity as:
\begin{equation}
\lim_{r_\rightarrow \infty}\kappa_{g}=\lim_{r_\rightarrow \infty}\frac{1}{Er}=0.
\end{equation}
The geodesic curvature vanishes at infinity and it decays with the radial coordinate. Therefore, at infinity the geodesics are straight lines, in other words, at infinity we have euclidean space.

\subsubsection{Near horizon geometry}
The near horizon limit of the geodesic curvature \eqref{gaussianc:1} can be calculated for the metric \eqref{gmetr:1} and we get

\begin{equation}\label{mpsc:1}
\kappa_{g}=-\frac{\sqrt{g(r)}}{2E\sqrt{f(r)}}f'=-\kappa_{surf},
\end{equation}   
where $\kappa_{surf}$ is the surface gravity of the black hole metric \eqref{gmetr:1}. Note that for non-extremal black holes $\kappa_{surf}>0$ then the geodesic curvature in \eqref{mpsc:1} is negative, and therefore, the geodesic curvature \eqref{mpsc:1} has to start from negative values at the horizon until it reaches zero at infinity. Geodesics near the horizon are not straight lines anymore, but have negative geodesic curvature.

\section{Stability of LR's and TMCO's}\label{slt:1}
The stability of circular orbits can be studied by analysing  the sign of the Gaussian curvature of the Jacobi metric \eqref{jacF:1}, the geodesics where the Gaussian curvature $\mathcal{K}$ is positive are deemed to be stable otherwise they are unstable. The criteria has been developed for spherically simmetric spacetimes in \cite{Qiao:2022jlu}, where a comparision with the conventional effective geoesic potential is performed. An extension to  more general static spacetimes is presented in \cite{Qiao:2022hfv}, see also \cite{Cunha:2022nyw}.

. Since we are studying MPS we enforce the condition \eqref{Fhcond:1}, then the Gauss curvature can be written
\begin{equation}\label{Kmps}
K_{MPS}=\sqrt{\frac{g}{F}}k'_{g}.
\end{equation}  
The equation \eqref{Kmps} shows that the stability of a geodesics of the Jacobi metric  and therefore, the stability of the spacetime geodesics (geodesics defined in the MPS), is determined by the sign of the derivative of the geodesic curvature of the Jacobi metric. If the geodesic curvature is a monotonically increasing function in the radial coordinate the geodesic is stable, otherwise it is going to be unstable. \\

\section{Massive particle surfaces and partial umbilicity}\label{sec:5}

The existence of geodesics can be inferred from the values of $\kappa_g$. In particular, for the existence of circular geodesics it is required that $\kappa_g=0$. This condition is directly related to the master equation \eqref{mastereq:1} that defines a MPS, a surface that is partially umbilic. In this section we show that the master equation for the existence of a MPS, and therefore the partial umbilic condition, is encoded in the geodesic curvature of the Jacobi metric \eqref{jacF:1}. In other words, the partial umbilicity property becomes a total umbilicty property but in the Jacobi metric. Our result is general enough that allows us to find even the master equation for massive charged particles defined in the appendix \ref{app:1}.\\
From the expression for the geodesic curvature of the Jacobi metric defined in \eqref{jacF:1} we obtain a condition for the existence of circular geodesics in the Jacobi metric \eqref{jacF:1}: 

\begin{equation}\label{Fhcond:1}
h'F+F'h=0.
\end{equation} 

The condition \eqref{Fhcond:1} is one of our important results. The expression $F h=cte$ encodes the partial umbilicity condition for the massive particle surfaces. Note that a MPS is a timelike surface (Lorentzian) and the Jacobi metric is a Riemannian metric. As far as we know, the idea that the Jacobi metric inherits the information regarding the MPS was not known until now. Moreover, the  equation \eqref{Fhcond:1} works for any type of massive particle worldlines, in particular for charged massive particles. The only thing that we need to calculate for every case is the conformal factor $F(r)$ of the metric \eqref{jacF:1}.\\
For example, if we take $F$ as defined in \eqref{Fjac} and replace in the condition  \eqref{Fhcond:1} we get

\begin{equation}\label{georest:1}
h'f-hf'-\frac{m^2}{\mathcal{E}^2}h'f^2=0,
\end{equation}
hence
\begin{equation}
\frac{\mathcal{E}^2}{m^2}=\frac{h'f^2}{h'f-hf'}.
\end{equation}

The previous equation is the equation \eqref{massivecond}. In addition, if $m=0$ we obtain from \eqref{georest:1} the condition \eqref{photoncond},  a umbilic condition for the photon sphere.
We have shown that the results about the geometry  of the MPS and the photon spheres are encoded in our formalism, and therefore we can carry and study both cases at the same time, even much more difficult cases such as the massive charged surface can be studied in full generality. Let see how the previous considerations work for different spacetime metrics.

\section{Examples}

We have shown that the master equation \eqref{mastereq:1} that has to be satisfied by  a MPS can be deduced from the geodesic curvature of the Jacobi metric \eqref{jacF:1}. Using equation \eqref{jacF:1} we can deduce the equation for marginal stable orbits and from this we can determine the condition for the ISCO and its radius $r_{ISCO}$. Furthemore, we can determine the photon sphere radius $r_{PS}$. In this section we use the condition \eqref{Fhcond:1} for calculating the master equation for different spacetime metrics. The fist part corresponds to the study of  MPS built with the worldlines of neutral particles in spacetime metrics such that the conformal factor $F(r)$ is the same as \eqref{Fjac}. The second part is dedicated to the study of the MPS built with the trajectories of charged particles, then the conformal factor in \eqref{Fhcond:1} has to be modified. The energy $\mathcal{E}$ has to include a term with the charge of the particle.

\subsection{Massive particle Surfaces for neutral particles}

\subsubsection{Schwarzschild geometry}
The first case that we are going to analyze is the Schwarzschild metric. We use the expression \eqref{Fhcond:1} with $f=1-\frac{2M}{r}$ and $h(r)=r^2$, then we obtain \footnote{This expression is obtained when the Killing vector is $\kappa^{\alpha}\partial_{\alpha}=\partial_{t}$. If the Killing vector is $\kappa^{\alpha}\partial_{\alpha}=\partial_{\phi}$ the master equation becomes 
\begin{equation}\label{Lm:1}
\frac{L^2}{m^2}=\frac{Mr^2}{r-3M}.
\end{equation} }.

\begin{equation}\label{em:1}
\frac{\mathcal{E}^2}{m^2}=\frac{(r-2M)^2}{r-3M}.
\end{equation}

The master equation \eqref{em:1} shows that a massive particle surface is defined for $r=cte$. Moreover, from that same equation we can find the ISCO  using $d\mathcal{E}/dr=0$, the resulting expression is the equation \eqref{isco:3}. The radius of ISCO can be calculated and it is $r_{ISCO}=6M$. Additionaly, from the denominator of the equation \eqref{em:1} we obtain the radius of the photon surface $r_{PS}=3M$, the point where the master equation diverges \cite{Claudel:2000yi}. Similarly, at $r_{ISCO}$ the momentum is given by $L=m\sqrt{12}M$. 

\subsubsection{Reissner-Nordström geometry}

As we did with the Schwarzschild metric we can proceed with the Reissner-Nordström metric where $f=1-\frac{2M}{r}+\frac{Q^2}{r^2}.$ Then, using equation \eqref{Fhcond:1} we get

\begin{equation}\label{noq}
\frac{\mathcal{E}^2}{m^2}=\frac{(r(r-2M)+Q^2)^2}{r^2(r(r-3M)+2Q^2)}.
\end{equation}

The radius of the ISCO orbit is obtained by solving
\begin{equation}
M(r^3-6Mr^2)+9MQ^2r-4Q^4=0.
\end{equation} 
The radius of the photon sphere is 
\begin{equation}
r_{PS}=\frac{1}{2}(3 M + \sqrt{9 M^2-8 Q^2}).
\end{equation}
The results obtained are very well known. Now we are going to study metrics such that $h(r)\neq r^2$.
\subsubsection{Fisher-Janis-Newman-Winicur}
Another interesting metric is the FJNW metric \cite{Wyman:1981,Virbhadra:1997ie}. This metric can be written
\begin{equation}
ds^2=-f^{\alpha}dt^2+\frac{dr^2}{f^{\alpha}}+h(r)(d\theta^2+\sin^2(\theta)d\phi^2)
\end{equation}
where
\begin{equation}\label{fa}
f=1-\frac{2M}{\alpha r},\,\,\,\,\alpha=\frac{M}{\sqrt{M^2+\Sigma^2}},\,\,\,\,h(r)=f^{1-\alpha}r^2.
\end{equation}
Using equation \eqref{georest:1} with $F$ defined in \eqref{jacF:1} and equations \eqref{fa} we obtain
\begin{equation}\label{fjnw}
\frac{\mathcal{E}^2}{m^2}= \left(1 - \frac{2 M}{\alpha r}\right)^\alpha \frac{(1 + \alpha) M - \alpha r}{(1+ 2 \alpha) M - \alpha r}.
\end{equation}

This expression is exactly the same expression obtained by the MPS method, see equation (66) in \cite{Kobialko:2022uzj}. 
By deriving $\mathcal{E}$ with respect to the radial coordinate in equation \eqref{fjnw} we can find the radius of the ISCO:
\begin{equation}\label{iscofjnw}
r_{ISCO \pm}=M\left(3 + \frac{1}{\alpha}\left(1 \pm\sqrt{5 \alpha^2-1}\right)\right),\,\,\,\, \alpha > \frac{1}{\sqrt{5}}
\end{equation}

By setting the denominator of the equation \eqref{fjnw} to zero we obtain
\begin{equation}\label{pscofjnw}
r_{PS}=\frac{(1+2\alpha)M}{\alpha},\,\,\,\,   1/2\leq\alpha\leq 1
\end{equation}

Both restrictions for $\alpha$ in equations \eqref{iscofjnw} and \eqref{pscofjnw} lead to four possible cases, see \cite{Kobialko:2022uzj}  for more details. When $\alpha=1$ we recover the Schwarzschild case. 

\subsubsection{Schwarzschild solution in Conformal gravity}

The Schwarzschild metric in conformal gravity is a spacetime constructed with the Schwarzschild metric multiplied by the conformal factor $\Omega=\left(1-\frac{l^4}{r^4}\right)$, thus now $f=\Omega^2 \left(1-\frac{2M}{r}\right)$ and $h(r)=\Omega^2 r^2$, we replace it in equation  \eqref{georest:1}  then we obtain
\begin{equation}
\frac{\mathcal{E}^2}{m^2}=\frac{(r-2M)^2}{r(r-3M)}\left(1-\frac{l^4}{r^4}\right).
\end{equation}
As before the ISCO condition reads
\begin{equation}
(4r^2-21 M r+30 M^2)l^4+r^4(r-6M)M=0                     
\end{equation}
The photon sphere radius is given by
\begin{equation}
r_{PS}=3M.
\end{equation}
As we have seen, all the cases we have presented are in accord with the result obtained by the MPS method. However, our approach is direct avoiding complicated geometrical calculations. Now we are going  to consider the motion of charged massive particles.

\subsection{Massive particle surfaces for charged particles}

\subsubsection{Charged particle in Reissner-Nordström geometry}

We consider the motion of particles with electrical charge $q$ and mass $m$ moving in spacetime metrics with charge. The conformal factor $F$ defined  is going to change and therefore we can not use equation \eqref{georest:1}, we need a more general expression.  We have to use the general condition \eqref{Fhcond:1}. The Jacobi metric for a charged particle in the Reissner-Nordström geometry has been calculated in \cite{Das:2016opi}, where the conformal factor $F$ is given by 
\begin{equation}\label{Fchrn}
F=\left(\frac{\left(\mathcal{E}-\frac{m q Q}{r}\right)^2-m^2 f(r)}{f(r)}\right),
\end{equation}
with $f(r)$ corresponding to the Reissner-Nordström metric. Note how the expression for the energy has changed because of the factor 
$\frac{m q Q}{r}$. Now, replacing equation \eqref{Fchrn} in the condition \eqref{Fhcond:1} we get

\begin{equation}\label{RNmaster}
\frac{\mathcal{E}_{\pm}}{m}=\frac{q
   Q r \left(r (r-4 M)+3 Q^2\right)\pm\left(r (r-2 M)+Q^2\right)^2 \sqrt{\frac{4 r (r-3 M)+\left(q^2+8\right) Q^2 r^2}{\left(r (r-2 M)+Q^2\right)^2}}}{2 r^2 \left(r (r-3 M)+2 Q^2\right)}.
\end{equation}

Here we can see how powerful our results are. The master equation \eqref{RNmaster} can be obtained using the massive particle surface presented in the appendix \ref{app:1}. The expressions described in the appendix \ref{app:1} are involved and difficult to calculate. Our new equation is in accordance with previous known results, see equation (57) in \cite{Das:2016opi}. The radius of the ISCO is a complicated expression, but it can be found numerically. For the photon sphere radius we get the expected result.  \\
Our approach is very simple and the only information that we need is the Jacobi metric, which was obtained by a projection of the spacetime metric over surfaces of constant energy. Moreover, our method allows to carry all the results, including the massless and the charged cases, in the same expression. Thus, when $q=0$ the expression in \eqref{noq} is recovered, and when $q=0$, $Q=0$ and $m=0$ we recover the photon sphere of the Schwarzschild case. 

\subsubsection{Electrically charged dilatonic black holes}
Another metric that we are going to analyze is the black hole in Einstein -Maxwell dilation theory, the so called electrically charged dilatonic black hole. The metric is given by \cite{Heydari-Fard:2021pjc}

 \begin{equation}\label{mecdbh}
 f(r)=\left(1-\frac{r_{+}}{r}\right)\left(1-\frac{r_{-}}{r}\right)^{\frac{1-a^2}{1+a^2}},\,\,\,\,\,h(r)=r^2\left(1-\frac{r_{-}}{r}\right)^{\frac{1-a^2}{1+a^2}}.
 \end{equation}
 
The conformal factor of the Jacobi metric becomes
\begin{equation}\label{Fecdbh}
F=\left(\frac{\left(\mathcal{E}-\frac{m q Q e^{2 a \phi_{\infty}}}{r}\right)^2-m^2 f(r)}{f(r)}\right).
\end{equation}

Replacing \eqref{Fecdbh} and \eqref{mecdbh} into \eqref{Fhcond:1} we obtain

\begin{equation}\label{ecdbh}
\frac{\mathcal{E}}{m}=\frac{\sqrt{Z}+e^{2 a} q Q r (r-r_{-})
   \left(\left(a^2+1\right) r (r-2 r_{+})+2 \left(a^2-1\right) r r_{-}-\left(a^2-3\right)r_{+}r_{-}\right)}{r^2 (r-r_{-}) \left(r\left(a^2+1\right)  (2 r-3 r_{+})+\left(a^2-3\right) r r_{-}+4 r_{+} r_{-}\right)},
\end{equation}

where

\begin{equation}
\begin{split}
Z &= m r^2 (r-r_{+})^2 (r-r_{-})(a^2+1)^2 e^{4 a} q^2 Q^2 (r-r_{-})^3\\
&+2 r^2 (a^2 r+r-r_{-}) \left(1-\frac{r_{-}}{r}\right)^{\frac{2}{a^2+1}} \left(\left(a^2+1\right) r (2 r-3 r_{+})+\left(a^2-3\right) r r_{-}+4 r_{+} r_{-}\right)
 \end{split}
\end{equation}

The photon sphere radius can be found by finding the points where the denominator of the expression \eqref{ecdbh} vanishes, namely
 \begin{equation}
\left(r\left(a^2+1\right)  (2 r-3 r_{+})+\left(a^2-3\right) r r_{-}+4 r_{+} r_{-}\right)=0.
 \end{equation}
The same results were found in \cite{Heydari-Fard:2021pjc} using the usual method, namely the geodesic approach. Later, these results were confirmed using the MPS method \cite{Bogush:2024fqj}. 

\section{Discussion}\label{discussion}

In this work we presented a study the Massive Particle Surfaces (MPS) using a Riemmanian approach. The MPS are timelike surfaces such that any worldline of a particle with mass, charge and fixed energy that is tangent to the MPS remains tangent all the time. There is a way to describe the geometry of MPS without using the worldlines of the particles. This leads to a modification  to the well known property of total umbilicity of the photon surfaces. \\
The partial umbilicity condition satisfied by MPS is the same condition that  photon surfaces satisfy in the Kerr spacetimes. The MPS correspond to  foliations of spacetime locally parameterized by energy values and the master equation that defines a MPS is an equation that relates the energy of the particle with an expression that needs to be constant.\\
We have provided a completely new characterization of the MPS using a Riemannian metric:the Jacobi metric. The master equation that characterizes partial umbilicity of a MSP can be deduced from the Jacobi metric which in its turn has been derived by projecting  a Lorentzian spacetime over a surface of constant energy. The condition \eqref{Fhcond:1} is one of our important results, and it encodes  the information of the master equation for the MPS \eqref{mastereq:1}. We only need the function $F$, which is the conformal factor of the Jacobi metric, plus the $h$ function which is the component $g_{\phi\phi}$ of the spacetime metric. The equation \eqref{Fhcond:1} is simpler and can be applied to any spacetime time metric. \\
The equation \eqref{Fhcond:1} is directly related with the existence of circular geodesics. Indeed, when the geodesic curvature is set to zero we obtain an equation that characterizes the circular geodesics, when $m=0$ our results reduce to the LR's cases. This geometric definition provides us with a condition for the partial umbilicity of the MPS. The expression is very simple and can be calculated for any metric of the type \eqref{gmetr:1}. Moreover, since we are using the geometry of the hypersurface built with the worldlines of the massive particles for characterizing the MSP, the condition \eqref{Fhcond:1} carries the information regarding the ISCO orbits and the information about the photon sphere. We have shown how this information can be extracted from different spacetime metrics.\\
There are certain limitations that need to be solved before applying the method to spacetimes with different asymptotic limits. The idea of characterizing the spacetimes  properties, such as horizons, is not well developed and we hope to give insights on this direction. Our results can be easily reproduced when the Killing vector is $\partial_{\phi}$. A direct generalization of the results found here would be to extend the formalism to horizonless spacetimes, such as wormholes \cite{Arganaraz:2019fup,Duenas-Vidal:2022kcx}. Finally, a theory characterizing the \textit{massive} shadows has been developed in \cite{Kobialko:2023qzo}, it would be interesting to characterize these shadows using a Riemannian geometric approach.

\appendix 
\section{Massive particle surfaces for Charged particles }\label{app:1}

Here we present a brief summary of the results concerning the massive particle surface approach for charged particles, we closely follow \cite{Kobialko:2022uzj}. Let us consider a stationary spacetime with a finite dimension $n$. This spacetime is endowed with a metric  that has a timelike Killing vector $k^{\mu}$ and a electromagnetic field described by the potential $A_{\alpha}$. We assume that both the metric and the vector potential share the same symmetry, namely $\mathcal{L}_{k}A_{\mu}=0$. We also consider the worldline $x^{\mu}$ of a charged massive particle, then its velocity $v^{\mu}=\dot{x}^{\mu}$ satisfies
\begin{equation}
v^{\mu}\nabla_{\mu}v^{\nu}=q F^{\nu}_{\sigma}v^{\sigma},\,\,\,\,v^{\mu}v_{\mu}=-m^2,
\end{equation}
where $F_{\mu\nu}=\partial_{\mu}A_{\nu}-\partial_{\nu}A_{\mu}$. A charged massive particle surface is a $n-1$ dimensional timelike surface $S$, with the property that if a particle with a fixed energy $\mathcal{E}=\mathcal{E}_{k}+\mathcal{E}_p$, where $\mathcal{E}_{k}=-k_{\mu}v^{\mu}$ and $\mathcal{E}_{p}=-qk_{\mu}{A^{\mu}}$, starts its motion in $S$ it will remain in $S$ forever. In order to have a charged massive particle surface, the right side of the following equation must be constant
\begin{equation}
\frac{\mathcal{E}^2_{\pm}}{m^2}=\sqrt{\frac{\kappa^2 \chi_{\tau}}{K}+\frac{\mathcal{F}^2(n-2)^2q^2}{4m^2K^2}}+\frac{\mathcal{F}(n-2)q}{2K}-qk_{\alpha}A^{\alpha},
\end{equation} 
where $k^2=\tilde{k}^{\mu}\tilde{k}_{\mu}$ and 
\begin{eqnarray}
K&=&-\chi_{\tau}+\frac{n-2}{2}n^{\mu}\nabla_{\mu}ln k^2\\
\mathcal{F}&=&\mathcal{F}^{\mu}_{\,\,\mu}=n^{\sigma}F_{\sigma\beta}\tilde{k}^{\beta},
\end{eqnarray}

An important aspect of the derivation is related with the partial umbilicity condition. The umbilicity property relates the induced metric, the extrinsic curvature and the electromagnetic field tensor:

\begin{equation}
\chi_{\alpha\beta}=\frac{\chi_{\tau}}{n-2}H_{\alpha\beta}+\frac{q}{\mathcal{E}_{k}}\mathcal{F}_{\alpha\beta},
\end{equation}

where
\begin{equation}
H_{\alpha\beta}=h_{\alpha\beta}+\frac{m^2}{\mathcal{E}^2_{k}}\tilde{k}_{\alpha}\tilde{k}_{\beta}
\end{equation}
and
\begin{equation}
h_{\alpha\beta}=g_{\alpha\beta}-n_{\alpha}n_{\beta},\,\,\,\chi_{\alpha\beta}=h^{\mu}_{\alpha}h^{\nu}_{\beta}\nabla_{\mu}n_{\nu}
\end{equation}

When $m=0$ and $q=0$ the partial umbilicity condition reduces to the known umbilicity condition for the photon surfaces.

\section*{Acknowledgments}
The work of BBC was funded by the National Agency for Research and Development (ANID)/ Scholarship Doctorado Nacional 2022/ folio 21220518.


\begin{thebibliography}{99}


\bibitem{Qiao:2022jlu}
C.~K.~Qiao and M.~Li,
``Geometric approach to circular photon orbits and black hole shadows,''
Phys. Rev. D \textbf{106} (2022) no.2, L021501,
\doi{10.1103/PhysRevD.106.L021501},
arXiv:\arxiv{2204.07297}[gr-qc].


\bibitem{Qiao:2022hfv}
C.~K.~Qiao,
``Curvatures, photon spheres, and black hole shadows,''
Phys. Rev. D \textbf{106} (2022) no.8, 084060,
\doi{10.1103/PhysRevD.106.084060},
arXiv:\arxiv{2208.01771}[gr-qc].

\bibitem{Cunha:2022nyw}
P.~Cunha, V.P., C.~A.~R.~Herdeiro and J.~P.~A.~Novo,
``Null and timelike circular orbits from equivalent 2D metrics,''
Class. Quant. Grav. \textbf{39} (2022) no.22, 225007,
\doi{10.1088/1361-6382/ac987e},
arXiv:\arxiv{2207.14506}[gr-qc].

\bibitem{Gibbons:2015qja}
G.~W.~Gibbons,
``The Jacobi-metric for timelike geodesics in static spacetimes,''
Class. Quant. Grav. \textbf{33} (2016) no.2, 025004
\doi{10.1088/0264-9381/33/2/025004}
arXiv:\arxiv{1508.06755}[gr-qc].

\bibitem{Arganaraz:2019fup}
M.~Arga\~naraz and O.~Lasso Andino,
``Dynamics in wormhole spacetimes: a Jacobi metric approach,''
Class. Quant. Grav. \textbf{38} (2021) no.4, 045004
\doi{10.1088/1361-6382/abcf86},
arXiv:\arxiv{1906.11779}[gr-qc].

\bibitem{Duenas-Vidal:2022kcx}
\'A.~Duenas-Vidal and O.~Lasso Andino,
``The Jacobi metric approach for dynamical wormholes,''
Gen. Rel. Grav. \textbf{55} (2023) no.1, 9,
\doi{10.1007/s10714-022-03060-w}
arXiv:\arxiv{2212.14147}[gr-qc].

\bibitem{Das:2016opi}
P.~Das, R.~Sk and S.~Ghosh,
``Motion of charged particle in Reissner\textendash{}Nordstr\"om spacetime: a Jacobi-metric approach,''
Eur. Phys. J. C \textbf{77} (2017) no.11, 735,
\doi{10.1140/epjc/s10052-017-5295-6},
arXiv:\arxiv{1609.04577}[gr-qc].


\bibitem{Maraner:2019ftj}
P.~Maraner,
``On the Jacobi metric for a general Lagrangian system,''
J. Math. Phys. \textbf{60} (2019) no.11, 112901,
\doi{10.1063/1.5124142},
arXiv:\arxiv{1912.08053}[physics.class-ph].


\bibitem{Claudel:2000yi}
C.~M.~Claudel, K.~S.~Virbhadra and G.~F.~R.~Ellis,
``The Geometry of photon surfaces,''
J. Math. Phys. \textbf{42} (2001), 818-838,
\doi{10.1063/1.1308507},
arXiv:\arxiv{gr-qc/0005050}[gr-qc].

\bibitem{Virbhadra:2002ju}
K.~S.~Virbhadra and G.~F.~R.~Ellis,
``Gravitational lensing by naked singularities,''
Phys. Rev. D \textbf{65} (2002), 103004,
\doi{10.1103/PhysRevD.65.103004}


\bibitem{Senovilla:2011np}
J.~M.~M.~Senovilla,
``Umbilical-Type Surfaces in Spacetime'',
arXiv:\arxiv{1111.6910}[math.DG].

\bibitem{Okumura:1967}
M. Okumura. "Totally umbilical hypersurfaces of a locally product Riemannian manifold." Kodai Math. Sem. Rep. 19 (1) 35 - 42, 1967. \doi{10.2996/kmj/1138845339}

\bibitem{Bogush:2023ojz}
I.~Bogush, K.~Kobialko and D.~Gal'tsov,
``Massive particle surfaces,''
Phys. Rev. D \textbf{108} (2023) no.4, 044070,
\doi{10.1103/PhysRevD.108.044070}

\bibitem{Kobialko:2022uzj}
K.~Kobialko, I.~Bogush and D.~Gal'tsov,
``Geometry of massive particle surfaces,''
Phys. Rev. D \textbf{106} (2022) no.8, 084032,
\doi{10.1103/PhysRevD.106.084032}
arXiv:\arxiv{2208.02690}[gr-qc].

\bibitem{Bogush:2024fqj}
I.~Bogush, K.~Kobialko and D.~Gal'tsov,
``Constructing massive particles surfaces in static spacetimes,''
Eur. Phys. J. C \textbf{84} (2024) no.4, 387,
\doi{10.1140/epjc/s10052-024-12751-4},
arXiv:\arxiv{2402.03266}[gr-qc].


\bibitem{Cunha:2020azh}
P.~Cunha, V.P. and C.~A.~R.~Herdeiro,
``Stationary black holes and light rings,''
Phys. Rev. Lett. \textbf{124} (2020) no.18, 181101,
\doi{10.1103/PhysRevLett.124.181101},
arXiv:\arxiv{2003.06445}[gr-qc].

\bibitem{Cunha:2017qtt}
P.~V.~P.~Cunha, E.~Berti and C.~A.~R.~Herdeiro,
``Light-Ring Stability for Ultracompact Objects,''
Phys. Rev. Lett. \textbf{119} (2017) no.25, 251102,
\doi{10.1103/PhysRevLett.119.251102},
arXiv:\arxiv{1708.04211}[gr-qc].

\bibitem{Ghosh:2021txu}
R.~Ghosh and S.~Sarkar,
``Light rings of stationary spacetimes,''
Phys. Rev. D \textbf{104} (2021) no.4, 044019
\doi{10.1103/PhysRevD.104.044019},
arXiv:\arxiv{2107.07370}[gr-qc].


\bibitem{Arganaraz:2021fwu}
M.~A.~Arga\~naraz and O.~Lasso ~Andino,
``A Riemannian geometric approach for timelike and null spacetime geodesics,''
Gen. Rel. Grav. \textbf{56} (2024) no.10, 121
\doi{10.1007/s10714-024-03314-9},
arXiv:\arxiv{2112.10910} [gr-qc].

\bibitem{Kre:1991}
Kreyszig, E. Differential Geometry. New York: Dover, p. 131, 1991.

\bibitem{Wyman:1981}
M. ~Wyman, ``Static spherically symmetric scalar fields in general relativity'',
Phys. Rev. D \textbf{24} (1981) no. 4, 839,
\doi{10.1103/PhysRevD.24.839}

\bibitem{Virbhadra:1997ie}
K.~S.~Virbhadra,
``Janis-Newman-Winicour and Wyman solutions are the same,''
Int. J. Mod. Phys. A \textbf{12} (1997), 4831-4836
\doi{10.1142/S0217751X97002577},
arXiv:\arxiv{gr-qc/9701021}[gr-qc].


\bibitem{Heydari-Fard:2021pjc}
M.~Heydari-Fard, M.~Heydari-Fard and H.~R.~Sepangi,
``Null geodesics and shadow of hairy black holes in Einstein-Maxwell-dilaton gravity,''
Phys. Rev. D \textbf{105} (2022) no.12, 124009,
\doi{10.1103/PhysRevD.105.124009},
arXiv:\arxiv{2110.02713}[gr-qc].

\bibitem{Kobialko:2023qzo}
K.~Kobialko, I.~Bogush and D.~Gal'tsov,
``Black hole shadows of massive particles and photons in plasma,''
Phys. Rev. D \textbf{109} (2024) no.2, 024060,
\doi{10.1103/PhysRevD.109.024060},
arXiv:\arxiv{2312.07498}[gr-qc].

\bibitem{Berg:1988}
B. G. M. Berger,  Differential Geometry: Manifolds, Curves and Surfaces. (1998) p. 416

\bibitem{Bambi:2016wdn}
C.~Bambi, L.~Modesto and L.~Rachwa\l{},
``Spacetime completeness of non-singular black holes in conformal gravity,''
JCAP \textbf{05} (2017), 003,
\doi{10.1088/1475-7516/2017/05/003},
arXiv:\arxiv{1611.00865}[gr-qc].

\end{thebibliography}
\end{document}